# Critical Point for Maximum Likelihood Decoding of Linear Block Codes

Marc Fossorier

***Abstract*— In this letter, the SNR value at which the error performance curve of a soft decision maximum likelihood decoder reaches the slope corresponding to the code minimum distance is determined for a random code. Based on this value, referred to as the critical point, new insight about soft bounded distance decoding of random-like codes (and particularly Reed-Solomon codes) is provided.**

## I. Introduction

Bounded distance decoding (BDD) has long been used as a design criterion for suboptimum soft decision decoders. Early works in this area are the generalized minimum distance (GMD) and Chase decoders [1], [2]. The main justification of this criterion is the fact that BDD has the same error correction radius as maximum likelihood decoding (MLD) in Euclidean space. At practical word error rates (WERs), BDD allows to achieve near-MLD of short block codes with relatively low complexity. However, it was indicated in [3] that this criterion becomes inappropriate for decoding longer codes at such WERs.

In this letter, we evaluate the SNR value at which the error performance of MLD becomes dominated by the minimum distance term for random codes. This value is referred to as the critical point for MLD of the code. It has been long recognized that classical upper bounds such as the union bound (UB) or the tangential sphere bound (TSB) [4] rapidly become tight as the SNR increases. However for medium to long code lengths, it is shown that despite their tightness, these bounds become dominated by the minimum distance term at quite low SNR values in general[1]. Consequently at relatively high SNR values, the performance loss of BDD over MLD can become even greater than that observed from simulations at practical WERs. In particular, for high rate Reed Solomon (RS) codes over GF(256), assuming their weight distribution is well approximated by that of a random code of the same distance [5], the critical points correspond to WERs $10^{-31}$ and $10^{-61}$ for the (255,239) and (255,223) codes, respectively (note that these WERs are lower than that of many standards based on concatenated systems with RS outer codes). These results also indicate that an error floor (or flare) may occur around the critical point for MLD of many long enough good codes.

## II. Critical Point for a Random Code

We assume an $(n, k, d_H)$ random-like binary code of length $n$, dimension $k$, rate $R = k/n$ and minimum distance $d_H$ is used for error control of BPSK transmission over an AWGN channel. By random-like code, we consider a code for which each term of its weight distribution is well approximated by the corresponding coefficient of a random code. As a result, turbo codes and many other iteratively decodable codes do not satisfy this definition. On the other hand, the binary images of high rate RS codes do [5] and constitute our main motivation. Another class of random-like codes is that of binary primitive BCH codes [6].

The WER of MLD is upper bounded by the UB [7]

$$P_e \leq \sum_{i=d_H}^{n} A_i Q\left(\sqrt{2RiE_b/N_0}\right) \tag{1}$$

with $A_i = 2^{-(n-k)}\binom{n}{i}$ for a random-like code. Assuming the SNR value $E_b/N_0$ is large enough for (1) to tightly approximate $P_e$, we have

$$\begin{aligned} P_e &\approx \max_i \left\{ 2^{-(n-k)} \binom{n}{i} \left(4\pi RiE_b/N_0\right)^{-1/2} \mathrm{e}^{-RiE_b/N_0} \right\} \\ &\approx \max_i \left\{ 2^{-n(1-R-H(i/n))} \mathrm{e}^{-RiE_b/N_0} \right\} \end{aligned} \tag{2}$$

with $H(p) = -p\log_2 p - (1-p)\log_2(1-p)$. Defining

$$f(i) = 2^{-n(1-R-H(i/n))} \mathrm{e}^{-RiE_b/N_0}, \tag{3}$$

and expressing $\partial f(i)/\partial i = 0$, simple algebra shows that for a given value $E_b/N_0$, $f(i)$ takes a unique maximum at

$$i = n\left(\mathrm{e}^{RE_b/N_0} + 1\right)^{-1}. \tag{4}$$

The critical point for MLD is obtained by setting $i = d_H$ in (4), which gives

$$(E_b/N_0)_{crit} = 1/R\ \ln(n/d_H - 1). \tag{5}$$

From (2), the corresponding WER is well approximated by

$$P_{e,crit} = 2^{-n(1-R-H(d_H/n))} \left(\frac{d_H}{n-d_H}\right)^{d_H}. \tag{6}$$

Note that for a random code achieving the Gilbert-Varshamov (GV) distance $d_{GV} = nH^{-1}(1-R)$, $(E_b/N_0)_{crit} = 1/R\ln(1/H^{-1}(1-R)-1)$ and $P_{e,crit} = (H^{-1}(1-R)/(1-H^{-1}(1-R)))^{nH^{-1}(1-R)}$.

If (5)-(6) provide a simple estimate of the critical point, tighter values can indeed be found. To this end, one can numerically apply the proposed approach to the TSB.

Supported by the NSF under Grant CCR-04-30576.

The author is with the Department of Electrical Engineering, University of Hawaii, Honolulu, HI 96822 (e-mail: marc@spectra.eng.hawaii.edu)

[1] If the union bound on MLD is dominated by the minimum distance term, then it is tight in most cases; however as shown in this letter, the reverse (occasionally a "folk" result implicitly assumed) is incorrect over a large range of WERs for many classes of codes.

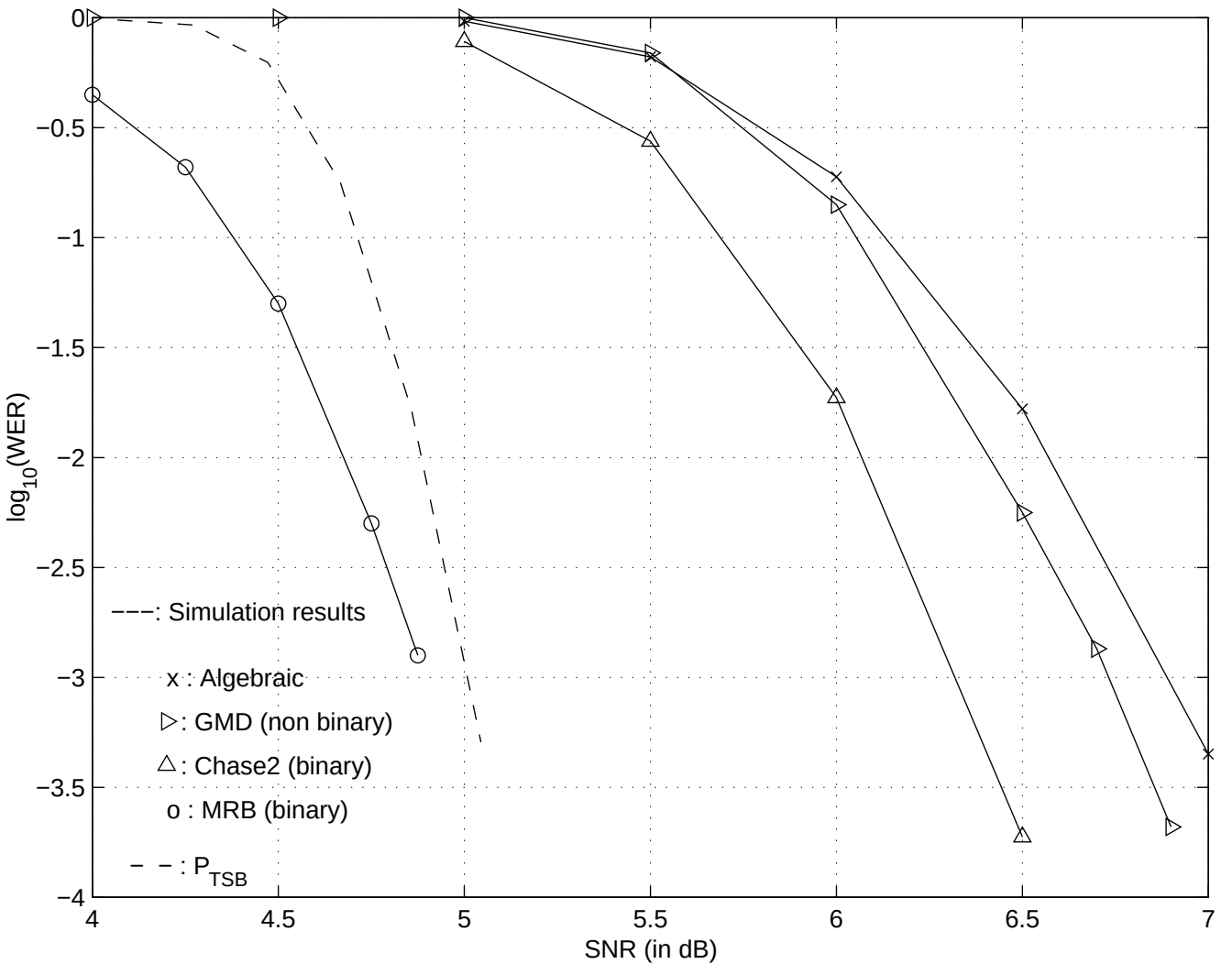

Fig. 1. WER for different decodings of the (255,239,17) RS code.

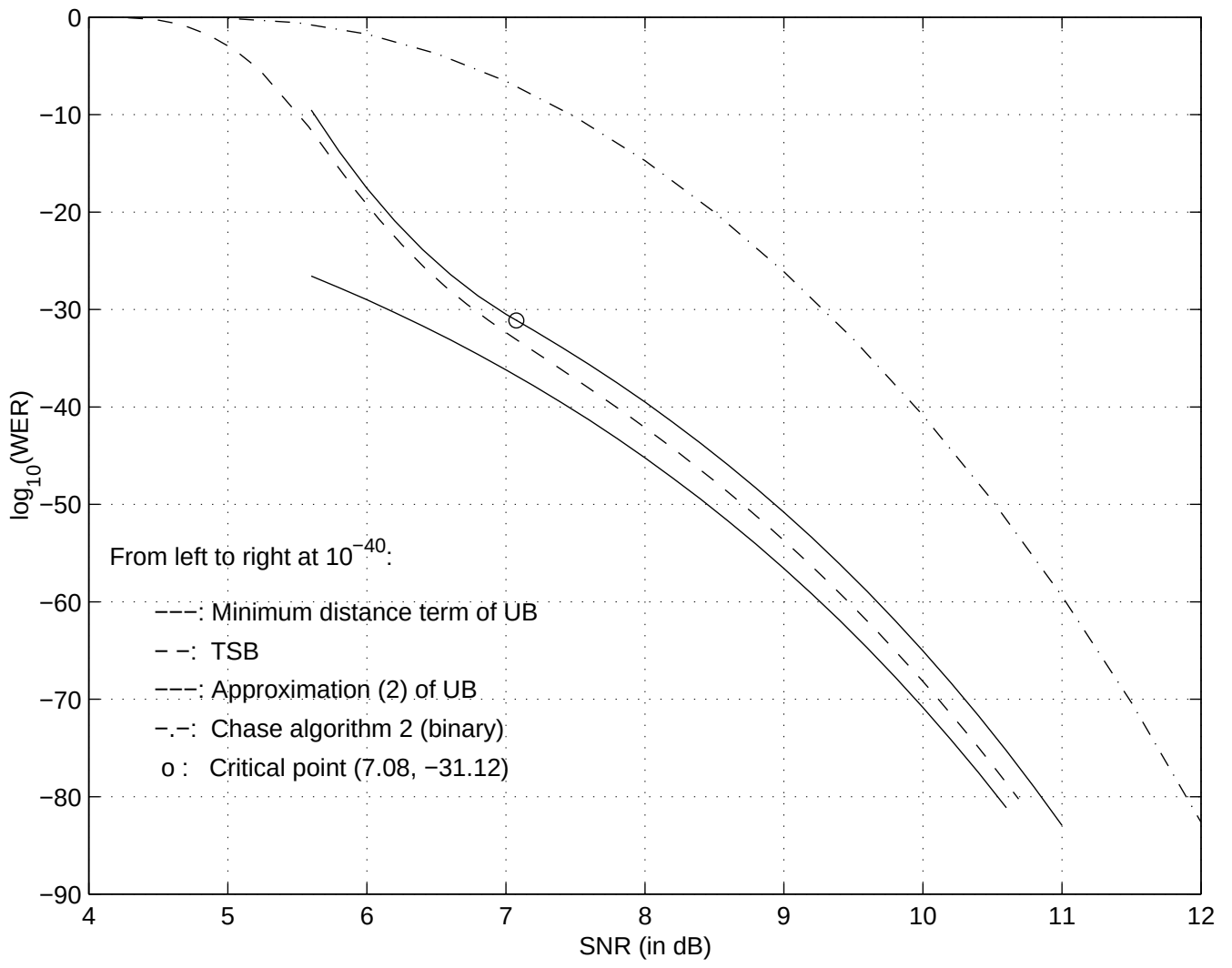

Fig. 2. Bounds and approximations for BDD of a (2040,1912,17) random-like code.

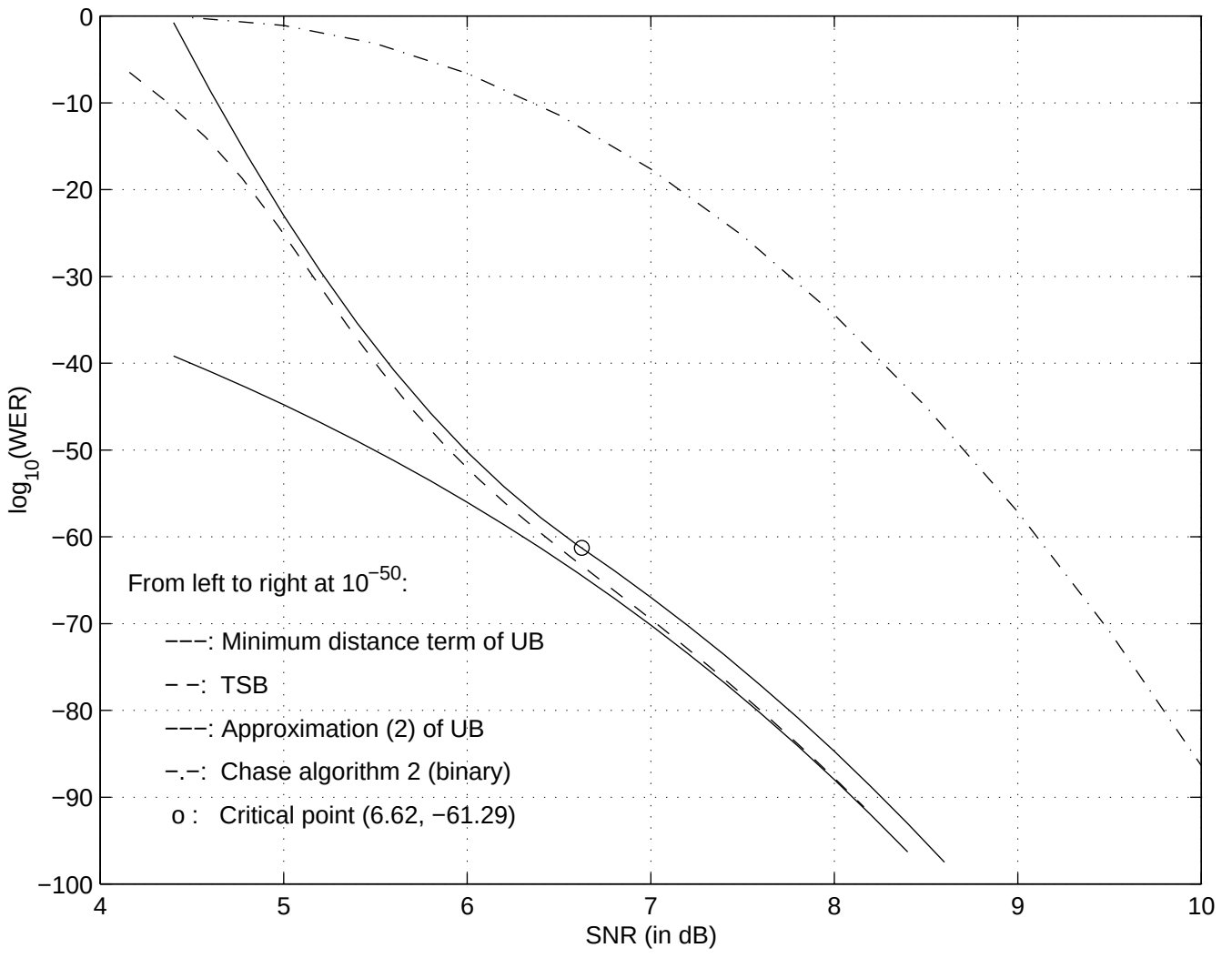

Fig. 3. Bounds and approximations for BDD of a (2040,1784,36) random-like code.

## III. Simulation Results

### A. RS(255,239) code

Fig. 1 depicts the WERs obtained for various decodings of the RS(255,239) code: algebraic BDD [7], GMD decoding at the non binary level as originally proposed in [1], Chase algorithm-2 decoding of the (2040,1912) binary image of the RS code [2] and the most reliable basis (MRB) reprocessing decoding of [8]. Applying the method proposed in [9], we verified that with a standard linear mapping, the binary image (2040,1912) has codewords of weight 17. Consequently all three soft decoding algorithms represented in Fig. 1 achieve BDD in Euclidean space. The TSB corresponding to the weight distribution of a (2040,1912) random code truncated at $d_H = 17$ has also been represented. We observe that the three BDD algorithms have very different performances. Furthermore, the slope of the TSB at these WERs is larger than that of any of the error performance curves represented, as suggested from the results of Section II.

In Fig. 2, the first term of (1) for $d_H = 17$, the approximation (2), the TSB, the tight bound on the WER of Chase algorithm-2 decoding obtained from order statistics [3], [10] and the critical point given by (5)-(6) have been depicted. We observe that the critical point accurately indicates a flaring of the MLD error performance curve. Importantly, the performance gap between the TSB and Chase algorithm-2 has increased from 1.45 dB at the WER $10^{-4}$ to 2.50 dB at the WER $10^{-30}$, despite the fact that Chase algorithm-2 achieves BDD. This gap remains larger than 1 dB at the WER $10^{-80}$. This figure confirms that the largest gain achieved by MLD over many BDD algorithms is likely to occur around the WER $10^{-30}$ for this code, as indicated by its critical point for MLD. Incidentally this WER is lower than that specified in most standards with this code.

### B. RS(255,223) code

The same curves as in Fig. 2 have been represented for a (2040,1784,36) random-like code in Fig. 3. This code has the same length and dimension as the binary image of RS(255,223), but the corresponding GV distance is $d_{GV} = 36$, against $d_H \geq 33$ for the binary image of RS(255,223). We observe that the critical point also indicates accurately a flaring of the MLD error performance curve around a much lower WER of $10^{-60}$.

### C. Random (2000,1000) code

Fig. 4 depicts the same curves as Fig. 2 and 3 for a random (2000,1000) code with minimum distance $d_{GV} = 222$. Indeed in that case, the Chase algorithm-2 is unfeasible but its error performance curve still represents an interesting reference. We observe that despite the fact that Chase algorithm-2 achieves BDD, the gap in error performance with respect to MLD remains about the same between the WERs $10^{-80}$ and $10^{-120}$. Comparing these three figures, we observe that for similar lengths, the critical point is reached at a lower WER as the rate decreases. Furthermore, the flaring is reduced as the rate decreases and in fact, almost no flaring is observed in Fig. 4.

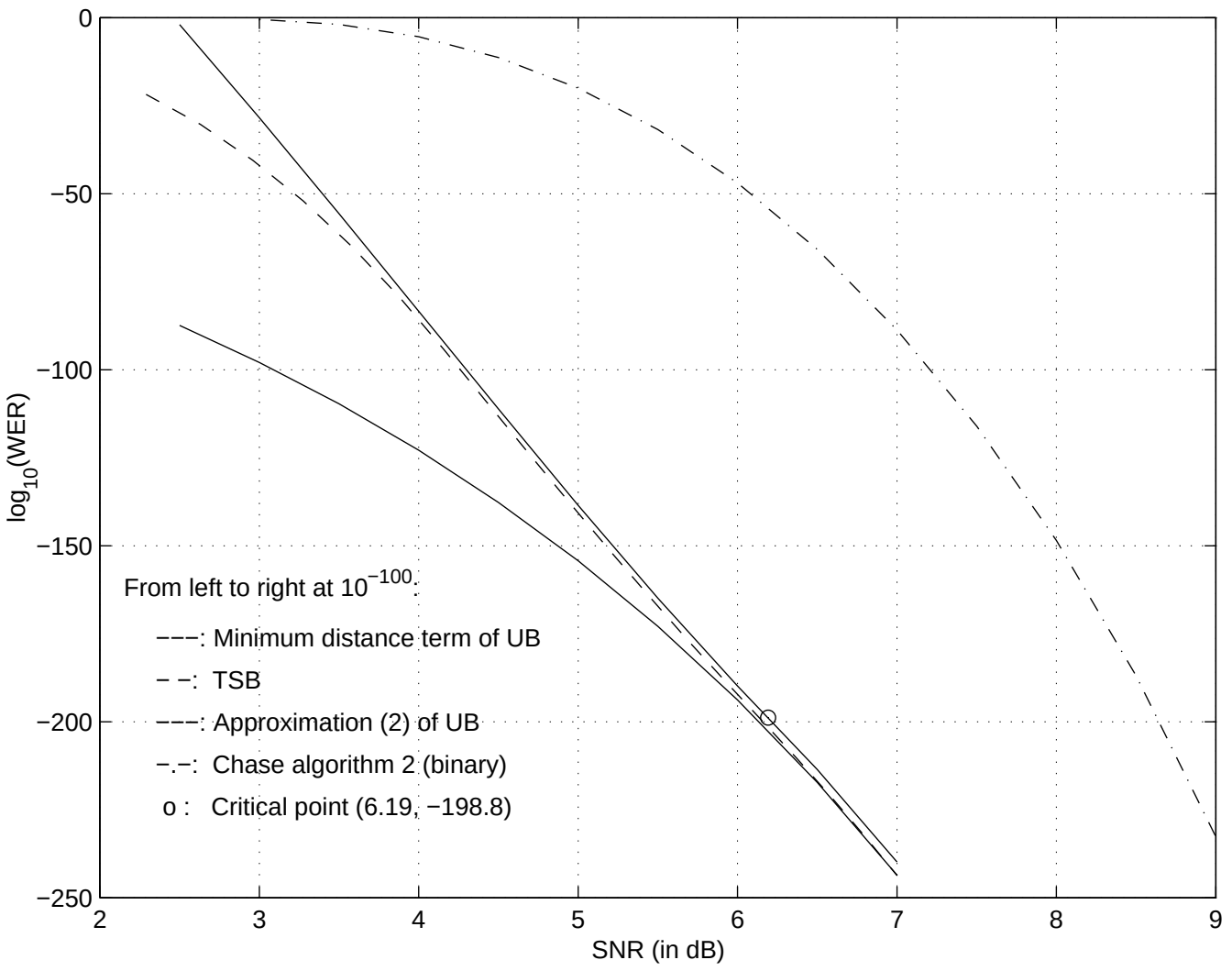

Fig. 4. Bounds and approximations for BDD of a (2000,1000,222) random-like code.

## IV. Conclusion

In this letter, a simple parameter referred to as the critical point for MLD has been introduced. This value indicates at which WER the error performance curve of MLD starts having the same asymptotic behavior as a BDD algorithm. It appears that for many long random-like codes, the critical point is located much below practical WER values.